\documentclass[aoas,MSNbibl,nameyear,dvips]{arximspdf}
\usepackage{graphicx}
\doi{10.1214/12-AOAS620} 
\volume{7}
\issue{2}
\pubyear{2013}
\firstpage{1074}
\lastpage{1094}

\makeatletter
\newcommand{\iid}{\stackrel{\mathrm{i.i.d.}}{\sim}}
\makeatother

\begin{document}
\begin{frontmatter}

\title{Bayesian nonparametric hierarchical modeling for multiple
membership data in grouped attendance interventions\thanksref{T1}}
\runtitle{Bayesian nonparametric modeling}
\thankstext{T1}{Supported by the National Institute on Alcohol Abuse
and Alcoholism grant to Susan Paddock (Grant Number R01AA019663). Data
collection was supported by the National Institute on Alcohol Abuse and
Alcoholism grant to Katherine Watkins (Grant Number R01AA014699).}

\begin{aug}
\author[A]{\fnms{Terrance D.} \snm{Savitsky}\corref{}\ead[label=e1]{tds151@gmail.com}}
\and
\author[A]{\fnms{Susan M.} \snm{Paddock}\ead[label=e2]{paddock@rand.org}}
\runauthor{T.~D. Savitsky and S.~M. Paddock}
\affiliation{RAND Corporation}
\address[A]{RAND Corporation\\
1776 Main Street, Box 2138\\
Santa Monica, California 90401-2138\\
USA\\
\printead{e1}\\
\phantom{E-mail:\ }\printead*{e2}}

\end{aug}

\received{\smonth{9} \syear{2012}}
\revised{\smonth{12} \syear{2012}}

%
\begin{abstract}
We develop a dependent Dirichlet process (DDP) model for repeated
measures multiple membership (MM) data. This data structure arises in
studies under which an intervention is delivered to each client through
a sequence of elements which overlap with those of other clients on
different occasions. Our interest concentrates on study designs for
which the overlaps of sequences occur for clients who receive an
intervention in a shared or grouped fashion whose memberships may
change over multiple treatment events. Our motivating application
focuses on evaluation of the effectiveness of a group therapy
intervention with treatment delivered through a sequence of cognitive
behavioral therapy session blocks, called modules. An open-enrollment
protocol permits entry of clients at the beginning of any new module in
a manner that may produce unique MM sequences across clients. We begin
with a model that composes an addition of client and multiple
membership module random effect terms, which are assumed independent.
Our MM DDP model relaxes the assumption of conditionally independent
client and module random effects by specifying a collection of random
distributions for the client effect parameters that are indexed by the
unique set of module attendances. We demonstrate how this construction
facilitates examining heterogeneity in the relative effectiveness of
group therapy modules over repeated measurement occasions.\vspace*{-3pt}
\end{abstract}

%
\begin{keyword}
\kwd{Bayesian hierarchical models}
\kwd{conditional autoregressive prior}
\kwd{Dirichlet process}
\kwd{group therapy}
\kwd{mental health}
\kwd{substance abuse treatment}
\kwd{growth curve}
\end{keyword}

\end{frontmatter}

\section{Introduction}\label{intro}

For many applications in which data have a multilevel structure,
observations on a study participant might not be nested within a single
higher level unit. Multiple membership (MM) modeling is used to account
for such data structures which arise in applications such as the
estimation of teacher effects from student test scores, where each
student is typically linked to multiple teachers over one or more
grades [\citet{HillGoldmult1998}]. MM structures also occur in the
analysis of health care costs when patients are treated by multiple
providers [\citet{Caremult2000}] and smoothing disease rates when
modeling health outcomes across geographic areas [\citet
{LangLeylRasbGoldmult1999}].

In our motivating application, the MM structure arises in a study of
the effect of group cognitive behavioral therapy (CBT) on reducing
depressive symptoms among clients in residential substance abuse
treatment. The Building Recovery by Improving Goals, Habits, and
Thoughts (BRIGHT) study [\citet{WatkHuntHepnPadddelaZhouGilm2011}]
was a community-based effectiveness trial of a group cognitive
behavioral therapy (CBT) intervention for treating residential
substance abuse treatment clients having depressive symptoms. The
BRIGHT study employed a quasi-experimental design in which cohorts of
clients at each of four study sites received either residential
treatment as usual (UC) ($n=159$) or residential treatment enhanced
with the BRIGHT intervention (CBT) provided by trained substance abuse
treatment counselors ($n=140$). Clients were assigned to receive either
CBT or UC according to which intervention was offered at their study
sites at the time of entry into residential substance abuse treatment.
CBT and UC were offered at each study site on an alternating basis over
time. The clients assigned to the CBT condition were expected to
complete four modules of group CBT, with each module consisting of four
thematically-similar sessions offered over a two-week period. This
sequence of modules was then offered on a repeating basis. In all,
$S=61$ group CBT modules were offered to the clients assigned to the
$\mathrm{CBT}$ condition. These~$61$ modules were divided into $G=4$ CBT
open-enrollment therapy groups, which are sequences of sessions that
have distinct sets of clients; the number of clients enrolled in each
open-enrollment group was $17$, $21$, $19$, and $83$, respectively.
Enrollment into the therapy group occurs on an open basis [\citet
{morgfals2006}, \citet{PaddHuntWatkMcCaAnal2011}], with clients entering
the therapy group at the start of new modules. The primary study
outcome is client depressive symptomology, as measured by the Beck
Depression Inventory-II (BDI-II) [\citet{becksteebrow1996}]. The
BDI-II score is a sum across $21$ four-level items (scored 0--3), with
a higher score indicating a greater level of depressive symptoms. The
BDI-II score for client $i$ is measured up to $o_i$ times, with $o_i=1$
for clients with only a baseline assessment at study entry and up to
$o_i=3$ for clients measured as well at both $3$ and $6$ months
post-baseline. The MM structure arises here since client outcomes might
be correlated due to common module attendance, and the BDI-II scores
are not uniquely associated with a single module but rather with all
modules attended by a client.

For longitudinal studies in which participants belong to multiple
higher-level units, the standard analytic approach is to include a
single set of random effects terms that are assumed to be constant over
time to account for the multiple membership. However, constraining
these random effects to be constant across time does not allow for
changes in correlations among outcomes for clients who attend modules
together; their outcomes might be more strongly correlated immediately
following group therapy versus at baseline or longer term follow-up
times. Further, including distinct terms in the model to account for
multiple membership and for the correlation of repeated measurements
within-client might be too restrictive for applications such as group
cognitive behavioral therapy (CBT). Not all clients benefit similarly
from group therapy [\citet{SmokRoseBacadama2001}]. For example, group
climate and cohesion are associated with improved outcomes [\citet
{RyumHageNordVogeStilgrou2009}, \citet{CrowGrenther2008}]. Thus, not
only might the effects of modules change over time, but also the
effects of modules on participant outcome trajectories might vary
across study participants.

We present a dependent Dirichlet process (DDP) model for repeated
measures multiple membership data. Specifically, we propose a set of
random distributions for client random effect parameters that are
indexed by therapy group module attendance sequences. Our model allows
one to obtain treatment effect estimates for group therapy versus a
comparison condition that account for the correlation of client
outcomes due to the attendance sequences, with the framework embedded
in a hierarchical construction for modeling repeated measures data. One
may use our approach to examine whether there is heterogeneity in the
relative effectiveness of group therapy modules by identifying clusters
of clients whose outcome trajectories vary across modules. Our
framework is flexible enough to retain application-specific modeling
choices. For the BRIGHT study, this includes specifying a proper
conditionally autoregressive (CAR) base distribution for the
nonparametric prior on module random effects, which accounts for the
open enrollment-induced client overlap in attendance of modules that
are offered at adjacent time points [\citet
{PaddHuntWatkMcCaAnal2011}]. We demonstrate that the DDP model may
be recast for estimation as a DP under our multiple membership linkage
of clients to treatment in a similar fashion as for the analysis of
variance (ANOVA) DDP [\citet{DeIMlleRosnMacEan2004}].

In Section \ref{addsect} we introduce an additive model that employs
client and MM random effects for BRIGHT study modules that was examined
for open-enrollment group therapy data by \citet{PaddSaviJRSSA}, and
then build upon that work by introducing a multivariate generalization
to allow for time-varying MM random effects. We present the DDP model
in Section \ref{ddpsect} to generalize the additive MM model to jointly
model dependence owing to repeated measures within clients and group
therapy module participation. Brief mention is made of our
computational approach and software solution for conducting posterior
simulations under the multiple membership models in Section \ref
{compsect}, followed by an exploration of the properties of the models
on simulated data in Section \ref{simsect}. Our motivating application
focuses on the assessment of a group CBT intervention deployed in an
open-enrollment study design for the treatment of depressive symptoms
among clients in residential substance abuse treatment in Section \ref
{case}. We conclude with discussion and conclusions in Section \ref
{discusssect}.

\section{Multiple membership additive semi-parametric models} \label{addsect}
This section introduces model constructions that include module random
effects, which are mapped to each client according to the modules
attended by that client using multiple membership modeling. These
models permit inference about the relative effectiveness of the CBT
intervention while accounting for differences in module effects as well
as the dependence induced among clients based on overlaps in the
sequences of modules attended. A separate client random effects term
captures the within-client dependence among repeated measures.
\subsection{Model construction and definitions}
We first begin with the model of \citet{PaddSaviJRSSA} for modeling
longitudinal post-treatment outcomes and allowing outcomes for clients
who attend the same therapy group to be correlated:
%
%
\begin{equation}
y_{ij} = \mu+ \mathbf{d}_{ij}^{\prime}\bolds{\beta} +
\mathbf{z}_{ij}^{\prime}\mathbf{b}_{i} +
\mathbf{x}_{i}^{\prime}\bolds{\gamma} + \varepsilon_{ij},
\label{modsess}
\end{equation}
 where $y_{ij}$ is the BDI-II depressive symptom score for
client $i$ ($i=1,\ldots,n$) at repeated measurement event $j =
(1,\ldots,o_{i})$. The global intercept is represented by~$\mu$. $\mathbf
{d}_{ij}$ are the fixed effects predictors and their associated effects
are $\bolds{\beta}$. We parameterize $\mathbf{d}_{ij} = (T_{i},
t_{ij}, t^2_{ij},T_{i} t_{ij}, T_{i} t^2_{ij} )$ for the BRIGHT
study, where $T_{i}$ specifies an indicator for the treatment arm
assigned to client $i$ [$T_{i}=1$ for clients receiving cognitive
behavioral therapy (CBT), $T_{i}=0$ for those receiving the ``usual
care'' (UC)] and $t_{ij}$ denotes the continuously-valued time at which
$y_{ij}$ was observed. The components of $\mathbf{d}_{ij}$ are chosen
to estimate the effects on depressive symptom scores of CBT assignment,
time, and the interaction of CBT assignment and time; a quadratic
specification was chosen based on previous data analysis [\citet
{PaddSaviJRSSA}]. The random effects predictor, $\mathbf{z}_{ij}$, is
a $q \times1$ vector associated with the $q$ random effects for client
$i$, $\{\mathbf{b}_{i}\}$. We set $\mathbf{z}_{ij} =
(1,t_{ij},t_{ij}^2)$ for the BRIGHT study, so that the $(q = 3) \times
1$ vector of random effect parameters for client $i$, $\mathbf{b}_{i}$,
capture client-specific variation in change in BDI-II scores over time.
Our parameterization of fixed and client random effects employs global
second order polynomial terms to enforce smoothness and prevent
overfitting under a study design with a relatively small number of
measurement waves per client, as is typical of behavioral intervention
studies such as BRIGHT. The second-to-last term allows for multiple
membership modeling since depressive symptom scores observed
post-treatment, $y_{ij}$, are not linked to specific therapy group
modules, but rather to all modules attended by client $i$. This term
maps the $y_{ij}$'s to the vector of~$S$ module random effects, $\bolds
{\gamma}$, by multiplying $\bolds{\gamma}$ by an $S \times1$ weight
vector, $\mathbf{x}_{i}$, that is normalized to sum to $1$ [\citet
{HillGoldmult1998}]. In particular, $S_i$ equals the number of
modules attended by client $i$; $x_{is}=1/S_i$ if client $i$ attended
module~$s$ and $x_{is}=0$ otherwise. Let $  N = \sum_{i} o_{i}$ denote
the number of repeated measures observed for all clients. Observational
error is indicated by $\varepsilon_{ij} \iid\mathcal{N} (0,\tau
_{\varepsilon}^{-1} )$. We produce within-sample fitted client growth
trajectories in Sections~\ref{simsect} and~\ref{case} with
employment of $ (\bolds{\beta},\{\mathbf{b}_{i}\},\bolds{\gamma} )$.

\subsection{Distribution of client random effects}
Though one may parametrically model the client random effects, $\{
\mathbf{b}_{i}\}$, we model them nonparametrically using a Dirichlet
process (DP) prior to motivate the subsequent DDP development and to
exploit the DP's usefulness for flexibly modeling the distribution of
the $\{\mathbf{b}_{i}\}$'s despite having no more than three repeated
measures per client in the BRIGHT study [\citet{PaddSaviJRSSA}]. Specify
%
%
\begin{eqnarray}
\label{dp} \mathbf{b}_{1},\ldots,\mathbf{b}_{n} \vert F &
\iid& F,
\\
F \vert\alpha, \bolds{\Lambda} &\sim& \operatorname{DP} (\alpha,F_{0} ),
\end{eqnarray}
where we choose the base distribution, $F_{0} \equiv\mathcal
{N}_{q} (\mathbf{0},\bolds{\Lambda}^{-1} )$, a convenient
conjugate form that spans the support for $\mathbf{b}$ and simplifies
posterior sampling while still allowing the data to estimate a general
form for $F$. We further specify $\alpha\sim\mathcal{G}a (a_{1} =
1, b_{1} = 1 )$ to allow the data to estimate the DP concentration
parameter, reflecting its importance for determining the total number
of client clusters formed. We may equivalently enumerate (\ref{dp}) as
a discrete mixture [\citet{sethuraman1994}],
%
%
\begin{equation}
\label{stick} F = \mathop{\sum}_{h=1}^{\infty}
p_{h}\delta_{\mathbf{b}^{\ast}_{h}},
\end{equation}
of countably infinite weighted point masses, where ``locations'' $
(\mathbf{b}^{\ast}_{1},\ldots,\mathbf{b}^{\ast}_{M} )$ index the
unique values for the $\{\mathbf{b}_{i}\}$. The discrete construction
for $F$ allows for ties among sampled values for $\{\mathbf{b}_{i}\}$,
so that $M \leq n$ and index clusters (i.e., clients sharing locations
or having same values of $\mathbf{b}$) with $n \times1,  \mathbf{s}$
where $s_{i} = m$ implies $\mathbf{b}_{i} = \mathbf{b}^{\ast}_{m}$.
Then the set, $ (\mathbf{s},\{\mathbf{b}^{\ast}_{m}\} )$,
provides an equivalent parameterization to $\{\mathbf{b}_{i}\}$, though
the former provides better mixing under posterior sampling [\citet{neal2000}].

\subsection{Distribution of module random effects} \label{modraneff}
\subsubsection{Univariate module effects}
Owing to the overlap in client attendance of modules under open
enrollment into group therapy,\vadjust{\goodbreak} we specify a conditionally
autoregressive (CAR) prior for module random effects to allow them to
be correlated. The degree of correlation is determined by the closeness
of the modules, which depends on how we define which modules are
neighbors. We define modules offered at adjacent time points within the
same open-enrollment group as neighbors given that clients tend to
attend subsequent modules in the BRIGHT study's residential treatment
setting [\citet{PaddHuntWatkMcCaAnal2011}].

To implement this, we enumerate a two-part form for the covariance
matrix [\citet{BesaMollYorkMollbaye1991}]. First, define an $S
\times S$ adjacency matrix, $\bolds{\Omega}$, to encode dependence
among neighboring modules where we set $\omega_{ss^{\prime}} = 1$ if module~$s$
is a \emph{neighbor} of or ``communicates'' with module~$s^{\prime}$
(denoted with ``$\sim$'' in $s \sim s^{\prime}$), and $0$ otherwise.
Construct $\mathbf{D} = \operatorname{Diag}(\omega_{s+})$, where
$\omega_{s+} =
\sum_{j} \omega_{sj}$ equals the number of neighbors of module $s$.
Then compose the covariance matrix, $ \mathbf{Q}^{-} = (\mathbf{D}
- \bolds{\Omega} )^{-}$, the Moore--Penrose pseudo-inverse, as $Q$
is not of full rank, and specify the joint distribution of random
module effects,
%
%
\begin{equation}
\label{eqcarprior} \bolds{\gamma}| \tau_{\gamma},\bolds{\Omega}
\sim
\mathcal{N} \bigl(\mathbf{0}, \bigl[\tau_{\gamma} (\mathbf{D} -
\bolds{
\Omega} ) \bigr]^{-} \bigr),
\end{equation}
where scalar precision parameter, $\tau_{\gamma}$, controls the overall
strength of variation. The rank of $ (\mathbf{D} - \bolds{\Omega
} )$ is $S - G$, where $G$ represents the number of distinct
open-enrollment therapy groups [\citet{HodgCarlFanon2003}].

We use the following model short-hand label for simulated data and
BRIGHT data analysis:
\begin{itemize}
\item\textit{MMCAR}: Employ the additive model of equation (\ref{modsess})
under the joint prior construction of equation (\ref{eqcarprior}) in the
fashion of \citet{PaddSaviJRSSA}.
\end{itemize}
Note that one could use a standard MM model for applications
under which random effects may be assumed exchangeable.

\subsubsection{Multivariate module effects}
The univariate module effects may be replaced with a multivariate model
specification that relaxes the assumption of constant module effects
over time specified in equation (\ref{eqcarprior}). Restate
equation (\ref
{modsess}),
%
%
\begin{equation}
y_{ij} = \mu+ \mathbf{d}_{ij}^{\prime}\bolds{\beta} +
\mathbf{z}_{ij}^{\prime}\mathbf{b}_{i} + \bigl(
\mathbf{x}_{i}^{\prime}\bolds{\Gamma} \bigr)\mathbf{z}_{ij}
+ \varepsilon_{ij}, \label{modmv}
\end{equation}
where $S \times q,  \bolds{\Gamma} = (\bolds{\gamma}_{1},\ldots,\bolds
{\gamma}_{S} )^{\prime}$, for each of the multivariate $q \times1$,
$\bolds{\gamma}_{s}$. We again assume a second order polynomial model, but
this time for the module effects, where each module, $s$, is
parameterized with a $(q=3) \times1$ random effects vector back
multiplied by $\mathbf{z}_{ij} = (1,t_{ij},t_{ij}^2)$, which permits
the effect of module $s$ under the BRIGHT study to vary with time,
$t_{ij}$. We may most easily make the extension of the CAR modeling of
\citet{BesaMollYorkMollbaye1991} by stacking each of the $q, S
\times1$ columns from $\bolds{\Gamma}$ into $qS \times1,
 \mathfrak
{G} = (\bolds{\gamma}_{(1)},\ldots,\bolds{\gamma}_{(q)} )$ for the
$S \times1,   \bolds{\gamma}_{(s)}$. Then compose the multivariate CAR prior,
%
%
\begin{equation}
\label{mcar} \mathfrak{G}|\bolds{\Lambda},\bolds{\Omega} \sim
\mathcal{N}
\bigl(\mathbf{0}, \bigl[ (\mathbf{D} - \bolds{\Omega} ) \otimes
\bolds{\Lambda}
\bigr]^{-} \bigr)
\end{equation}
for the $qS \times qS$ precision matrix, $\mathbf{Q} = (\mathbf{D}
- \bolds{\Omega} ) \otimes\bolds{\Lambda}$, where $\bolds
{\Lambda}$ describes the dependence among the $q$ random effects per
module and is specified to be identical to that used for the base
distribution associated with the prior (\ref{dp}) imposed on $\{
\mathbf
{b}_{i}\}$. In summary, equation (\ref{modmv}) extends equation (\ref
{modsess}) by permitting MM random (module) effects to vary over time.
Assign the following label for our multivariate construction:
\begin{itemize}
\item\textit{MM\_MV}: Employ the additive model of equation (\ref{modmv})
under the joint prior construction of equation (\ref{mcar}).
\end{itemize}

\subsection{Prior distributions for other parameters}
Scalar precision parameters $(\tau_{\varepsilon},\tau_{\gamma})$ are each
specified with a $\mathcal{G}a(0.1,0.1)$ prior with mean $1$, while the
$q \times q$ precision matrix $\bolds{\Lambda} \sim\mathcal
{W}(q+1,\mathbb{I}_{q})$, where the degrees of freedom are set to the
minimum value to encourage updating by the data. Last, $(\mu,\bolds
{\beta
})$ each receive noninformative priors. In instances where our priors
specify fixed hyperparameters, we use values intended to be easily
overwhelmed in the presence of data rather than eliciting them from our data.

\section{Dependent Dirichlet process for multiple membership data}
\label{ddpsect}
To allow for greater flexibility in modeling changes in module effects
over time as well as the effects of modules on client depressive
symptom trajectories, we now reformulate equation (\ref{modsess}) to
explicitly index the client random effects by group therapy module
identifiers, under which each client is assigned a $q \times(S+1)$
matrix of random effects. This contrasts with the previous
specification of sets of $q \times1$, $\{\mathbf{b}_{i}\}_{i=1,\ldots,n}$ client random effects and the $S \times1$ module effects, $\bolds
{\gamma}$, given in equation~(\ref{modsess}). The resulting
client-by-module matrix parameterization arises from replacing a single
random prior distribution for client effects with a collection of
random prior distributions that are indexed by the unique module
attendance sequences. First, we reformulate equation~(\ref{modsess}) in a
more flexible composition,
%
%
\begin{eqnarray}
y_{ij} &=& \mu+ \mathbf{d}_{ij}^{\prime}
\bolds{\beta} + \mathbf{z}_{ij}^{\prime}\bolds{\Delta}_{i}
\mathbf{x}_{i}+ \varepsilon_{ij},\label{interact}
\\
\bolds{\Delta}_{1},\ldots,\bolds{\Delta}_{n} \vert F &
\iid& F, \label{dpdelt}
\\
F \vert F_{0} &\sim& \operatorname{DP} (\alpha,F_{0} ),
\label{Fprior}
\end{eqnarray}
where we have replaced $q \times1, \mathbf{b}_{i}$ and $S \times
1, \bolds{\gamma}$ with the $q \times(S+1), \bolds{\Delta}_{i}$ for
client~$i$ composed with
%
%
\begin{equation}
\bolds{\Delta}_{i} = [\mathbf{b}_{i},
\mathbf{a}_{1,i},\ldots,\mathbf{a}_{S,i} ].
\end{equation}
The first column of $\bolds{\Delta}_{i}$ employs the analogous
$\mathbf
{b}_{i}$ client random effects from the additive models. The $\{\mathbf
{a}_{s,i}\}_{s = 1,\ldots,S}$ collect a set of $q \times1$ module
random effect vectors for client $i$. We note that every client
receives an effect term, $\mathbf{a}_{s,i}$, for all of the $S$
modules, even for modules they have not attended; such is even true for
clients in the UC arm. By contrast, the additive model of equation (\ref
{modsess}) is only defined at observed sequences of client module
attendances, while this formulation is defined over a broader space of
potential module attendance sequences across clients. We impose a DP
prior on the set of client-by-module effects, $\bolds{\Delta}_{i}$, in
order that we may borrow strength and dimension reduce to discover
clusters of clients expressing differential response sensitivities to
treatment exposures. Employment of a continuous base distribution under
the DP prior for the $\{\bolds{\Delta}_{i}\}_{i=1,\ldots,n}$ allows
the posterior inference on an arbitrary sequence of group therapy
module linkages for each client. Effect values at unobserved modules
are drawn from the nondegenerate continuous base distribution as
updated by the \emph{observed} module attendances. The module effect
estimates for unobserved attendances for each client are set equal to
the location values associated with the cluster to which the client is
assigned. The ability to develop a proper posterior distribution for
arbitrary module attendance sequences is referred to by \citet
{DeIMlleRosnMacEan2004} as nondegeneracy.

Each of the $q \times1$ columns of $\bolds{\Delta}_{i}$ in
equation (\ref{interact}) is back multiplied by $\mathbf{x}_{i}$, which
is the MM weight vector we earlier defined, but with a $1$ prepended
for a random intercept. More specifically, for $\mathbf{x}_{i}$ equal
to some value $\mathbf{x}$, we construct the latter object as $\mathbf
{x} \equiv(1, x_{1},\ldots,x_{S} )$ for $x_{s} \in
[0,1 ]$ to encode the vector sequence for group therapy module
attendance. Under our MM construction, the $(S+1) \times1, \mathbf{x}$
is composed of values in $[0,1]$ for $\sum_{s=1}^{S}x_{s} = 1$ for
clients who attend at least one module, and $\sum_{s=1}^{S}x_{s} = 0$
for clients who do not.

We define the $q \times1$ parameter vector, $\bolds{\theta}_{\mathbf
{x},i} \equiv\bolds{\Delta}_{i}\mathbf{x}$, resulting from
composition of the client-by-module random effects with the module
attendance sequence. We write $\bolds{\theta}_{\mathbf{x},i}$ and $\bolds
{\theta}_{\mathbf{x},i^{\prime}}$ for clients $i$ and $i^{\prime}$ that share the
same attendance sequence, $\mathbf{x} \in\mathcal{X}$. Construct the
subsequence, $(x_{s(1)},\ldots,x_{s(K)} )$ for $K \leq S$
nonzero entries in $\mathbf{x}$ corresponding to modules attended for
one or more clients with $\mathbf{x}_{i} = \mathbf{x}$. Then we may
provide the more granular construction, $\bolds{\theta}_{\mathbf{x},i} =
\mathbf{b}_{i} + x_{s(1)}\mathbf{a}_{s(1),i} + \cdots+
x_{s(K)}\mathbf
{a}_{s(K),i}$, for client $i$ where we note that only those modules
attended by client~$i$ contribute to the likelihood. The multiplication
of each $\mathbf{a}_{s(k)}$ by $x_{s(k)}$ reflects the MM design with
$x_{s(k)} \in[0,1]$.

Our formulation in equation (\ref{interact}) may be re-expressed with the
$q \times1$ vector of client random effects, $\bolds{\theta}_{\mathbf
{x},i}$, in a similar fashion as the $q \times1$ $\mathbf{b}_{i}$ in
equation (\ref{modsess}), but here we index the client random effects by
module attendance sequence $\mathbf{x}$. The prior for $\bolds{\theta
}_{\mathbf{x},i}$ is specified under a collection of random
distributions, $\{F_{\mathbf{x}}\}$, indexed by the unique attendance
sequences, $\mathbf{x} \in\mathcal{X}$,
%
%
\begin{equation}
y_{ij} = \mu+ \mathbf{d}_{ij}^{\prime}\bolds{\beta} +
\mathbf{z}_{ij}^{\prime}\bolds{\theta}_{\mathbf{x},i} +
\varepsilon_{ij}, \label{modddp}
\end{equation}
with random effects vector, $\mathbf{z}_{ij}$, the same as composed in
equation (\ref{modsess}). Specify the prior formulation for $\bolds{\theta
}_{\mathbf{x},i}$,
%
%
\begin{eqnarray}
\label{mmddp} \bolds{\theta}_{\mathbf{x},i} |F_{\mathbf{x}} &\iid&
F_{\mathbf{x}}.
\end{eqnarray}

We next enumerate a multiple membership dependent Dirichlet process (MM
DDP) set of nonparametric distributions indexed by the module
attendance sequence, $\mathbf{x}$, in the stick-breaking construction
[\citet{sethuraman1994}],
%
%
\begin{equation}
\label{stickddp} F_{\mathbf{x}} = \mathop{\sum}_{h=1}^{\infty}
p_{h}\delta_{\bolds{\theta
}^{\ast}_{\mathbf{x},h}},
\end{equation}
of weighted point mass locations where the weights are common for all
values of $\mathbf{x} \in\mathcal{X}$, but the locations are indexed
the unique attendance sequences (unlike for the simpler DP). We note
that marginally, for each $\mathbf{x}$, the locations $\bolds{\theta
}^{\ast
}_{\mathbf{x},h}$ are exchangeable in $h$, such that $F_{\mathbf{x}}$
follows a Dirichlet process and we have established the propriety of
the MM DDP. Denote the following short-hand notation for MM DDP construction,
%
%
\begin{eqnarray}
\label{mmddp} \bolds{\theta}_{\mathbf{x},i}|F_{\mathbf{x}} &\iid&
F_{\mathbf{x}},
\\
\{F_{\mathbf{x}}, \mathbf{x} \in\mathcal{X}\} &\sim& \operatorname
{MM\, DDP} (
\alpha,F_{0} ),
\end{eqnarray}
where we have extended the ANOVA DDP prior of \citet{DeIMlleRosnMacEan2004} to a
multiple membership framework for the set of effect random
distributions, $\{F_{\mathbf{x}}\}$.

We achieve equation (\ref{interact}) from equation (\ref{modddp}) by
extending a property of ANOVA DDP to the MM DDP that rewrites
equation (\ref{stickddp}) as a DP due to the finite indexing space of
group therapy modules with
%
%
\begin{eqnarray}
F_{\mathbf{x}} &=& \mathop{\sum}_{h=1}^{\infty}
p_{h}\delta_{\bolds{\theta
}^{\ast}_{\mathbf{x},h}}
\\
&=& \mathop{\sum}_{h=1}^{\infty} p_{h}
\delta_{\bolds{\Delta}^{\ast
}_{h}\mathbf{x}} = \delta_{\mathbf{x}}\mathop{\sum}_{h=1}^{\infty}
p_{h}\delta_{\bolds{\Delta}^{\ast}_{h}},
\\
F_{\mathbf{x}} &=& \delta_{\mathbf{x}}F.
\end{eqnarray}
Then we may rewrite our DDP model formulation of equation (\ref{modddp})
to the DP construction specified in equation (\ref{interact}).

Though we use equations (\ref{interact})--(\ref{Fprior}) to estimate the
MM DDP, the conceptual alternative in equations (\ref{modddp})--(\ref
{stickddp}) provides insight into the inferential properties of the MM
DDP. The indexing of distributions, rather than just mean effects, by
the module attendance sequences better spans the space of distributions
generating the client random effects and allows the estimation of
client module effects for modules not attended.

We also gain insight into the manner in which strength is borrowed over
the set of module attendance sequences. The MM DDP formulation employs
$\{F_{\mathbf{x}}\}_{\mathbf{x} \in\mathcal{X}}$ indexed by the set of
unique module attendance sequences. Few clients, however, may be
expected to exactly overlap or to share the same $\mathbf{x}$. Yet
clients will overlap for a portion of the module attendance sequences
such that we have repeated observations for each module $s \in
(1,\ldots
,S)$ for estimation of the dependent $\{\mathbf{a}_{s,i^{\prime}}\}_{i^{\prime}}$
for all $i^{\prime}\dvtx x_{s,i^{\prime}} > 0$. The partial overlaps among the $\{
\mathbf{x}\}_{\mathbf{x} \in\mathcal{X}}$ induce a dependence
structure among the $\{F_{\mathbf{x}}\}$ based on the extent of overlaps.

\subsection{Base distribution} \label{base}
We structure the base distribution, $F_{0}$, for our $q \times(S+1)$
client-by-module parameters to leverage the adjacency dependence of the
BRIGHT study modules. Compose $F_{0}$ for draws for the cluster
locations, $\{\bolds{\Delta}^{\ast}_{m}\}_{m = 1,\ldots,M}$, as the
product of multivariate Gaussian distributions for each of the $q
\times1$, $\mathbf{b}^{\ast}_{m}$ and the $q \times S$, $\mathbf
{A}^{\ast}_{m} = [\mathbf{a}_{1,m},\ldots,\mathbf{a}_{S,m} ]$
that, together, comprise $\bolds{\Delta}^{\ast}_{m} = [\mathbf
{b}^{\ast}_{m}, \mathbf{A}^{\ast}_{m} ]$ with
%
%
\begin{eqnarray}
\mathbf{b}^{\ast}_{m} \vert\bolds{\Lambda} &\iid& \mathcal
{N}_{q} \bigl(\mathbf{0},\bolds{\Lambda}^{-1} \bigr),
\\
\mathbf{A}^{\ast}_{m} \vert\bolds{\Lambda},\bolds{\Omega},
\rho&\iid& \mathbf{0} + \mathcal{N}_{q \times S} \bigl(\bolds{\Lambda
}^{-1},\mathbf{Q}^{-1} \bigr),\label{dawid}
\end{eqnarray}
where $m$ indexes cluster location. The $\mathcal{N}_{q \times S}$
construction in equation (\ref{dawid}) employs a separable (parsimonious)
covariance formulation for the distribution on the set of $q \times S$
matrix variate parameters, $\mathbf{A}^{\ast}_{m}$. We have employed
the notation of \citet{Dawisome1981} under which the $q \times q$,
$\bolds{\Lambda}$, defines the precision matrix for the columns of
$\{
\mathbf{A}^{\ast}_{m}\}$ and the $S \times S$, $\mathbf{Q}$, for the
rows. The covariance formulation is equivalent to $\operatorname{Cov}
[\operatorname
{vec} (\mathbf{A}^{\ast}_{m} ) ] = \bolds{\Lambda}^{-1}
\otimes\mathbf{Q}^{-1}$. [See \citet{Hoffsepa2011} for an intuitive
discussion of separable covariance formulations.] Last, the preceding
$\mathbf{0}$ presents the value of the $q \times S$ mean. Consistent
with prior formulations under the additive models, the $q \times
q, \bolds{\Lambda} \sim\mathcal{W}(q+1,\mathbb{I}_{q})$. We structure
the $S \times S$ precision matrix, $\mathbf{Q}$, which models the
module-induced adjacency dependence among the $q \times1$ set, $\{
\mathbf{a}^{\ast}_{m,s}\}_{s = 1,\ldots,S}$, with a proper CAR
formulation as enumerated in \citet{JinCarlBanegene2005}, where
$\mathbf{Q} = (\mathbf{D} - \rho\bolds{\Omega} )$ and $\rho
\in(-1,1)$ ensures $\mathbf{Q}$ is of full rank and may be viewed as a
smoothing parameter that measures the strength of the adjacency
association. Matrices $ (\mathbf{D},\bolds{\Omega} )$ hold
the same definitions as earlier specified in Section \ref{modraneff}.

Proceeding with the notation of \citet{Dawisome1981}, we pull
together the components of the base distribution into
%
%
\begin{equation}
F_{0} = f \bigl(\bolds{\Delta}^{\ast}_{m} \vert
\bolds{\Lambda},\bolds{\Omega},\rho\bigr) = \mathbf{0} +
\mathcal{N}_{q \times
(S+1)}
\bigl(\bolds{\Lambda}^{-1},\mathbf{P}^{-1} \bigr),
\end{equation}
where $\mathbf{P} = \operatorname{diag} (1,\mathbf{Q} )$. Let us
prepare $F_{0}$ in the form we will use to conduct posterior
simulations by stacking the $q$ rows of $\bolds{\Delta}^{\ast}_{m}$
[each an $(S+1) \times1$ vector] to the $q(S+1) \times1, \bolds{\delta
}^{\ast}_{m} = (\delta_{1,m}^{\prime},\ldots,\delta_{q,m}^{\prime}
)^{\prime}$ in
%
%
\begin{equation}
\label{F0} F_{0} = f \bigl(\bolds{\delta}^{\ast}_{m}
\vert\bolds{\Lambda},\bolds{\Omega},\rho\bigr) = \mathcal{N}_{q(S+1)}
\bigl(\mathbf{0}, [\bolds{\Lambda} \otimes\mathbf{P} ]^{-1}
\bigr).
\end{equation}

Vectorize $\mathbf{A}^{\ast}_{m}$ in a similar manner to obtain the $qS
\times1$, $ \mathfrak{a}^{\ast}_{m} \vert\bolds{\Lambda},\bolds{\Omega},\rho\iid\mathcal{N} (\mathbf{0}, [ (\mathbf{D}
- \rho\bolds{\Omega} ) \otimes\bolds{\Lambda} ]^{-1}
)$, which is similar to (\ref{mcar}) but is full rank to permit
efficient joint posterior sampling under high within-cluster dependence
among the $qS$ elements of $\mathfrak{a}^{\ast}_{m}$. Our MM DDP formulation specifies the full set of $S$ module
effects for client $i$ set equal to the location values, $\{\mathfrak
{a}^{\ast}_{m}\}$, drawn from the CAR base distribution for cluster $m$
that contains client $i$ for some posterior sampling iteration.

Due to the BRIGHT study design, there were $G=4$ open-enrollment
therapy groups. Each group was composed of modules having at least
partial overlap with another module with respect to the set of clients
in attendance, and the sets of clients in the four groups were
different. We thus add more flexibility in (\ref{F0}) by specializing
the CAR prior in $\mathbf{P}$ to each open-enrollment therapy group with
%
%
\begin{equation}
\mathbf{P} = \operatorname{diag} (1,\mathbf{Q}_{1},
\mathbf{Q}_{2},\ldots,\mathbf{Q}_{G} ),
\end{equation}
where we have defined a set, $\{\mathbf{Q}_{g}\}_{g=1,\ldots,G}$, of
CAR precision matrices composed as $S_{g} \times S_{g}, \mathbf{Q}_{g}
= (\mathbf{D}_{g} - \rho_{g}\bolds{\Omega}_{g} )$ and
recover $\mathbf{D} = \operatorname{diag} (\mathbf{D}_{1},\ldots,\mathbf
{D}_{G} )$ and $\bolds{\Omega} = \operatorname{diag} (\bolds
{\Omega
}_{1},\ldots,\bolds{\Omega}_{G} )$, reflecting the disjoint,
noncommunicating structure we seek to model. It is noted by \citet
{JinCarlBanegene2005} that the parameterization of the global
scalar smoothing parameter, $\rho$, may be overly restrictive, and they
offer more heavily parameterized alternatives to permit the learning to
adapt more locally. Our specification that offers the indexing of $\rho
_{g}$ by disjoint group allows smoothing across client-indexed module
effects to be local to group. We may specify other continuous,
multivariate distributions in place of the CAR for each group,
including replacing the CAR covariance matrix construction with an
anistropic Gaussian process \citet{savitsky2010} or with an
unspecified formulation under an inverse Wishart prior.

Assign the following label for the nonparametric construction:
\begin{itemize}
\item\textit{DDP}: Equations (\ref{modddp})--(\ref{stickddp}) under the base
distribution of equation (\ref{F0}).
\end{itemize}

\section{Computational approach} \label{compsect}
Convergence of the sampler employed for simulation and the BRIGHT data
analyses was assessed by employing a fixed width estimator with Monte
Carlo standard errors (MCSE) computed using the consistent batch means
(CBM) method [\citet{JoneHaraCaffNeatfixe2006}]. Computational
software for the posterior distribution simulations is available in our
package for the R statistical software [\citet{rdctr}] package called
\texttt{growcurves} [\citet{SaviPaddgrow2011}]. All of the methods, fit
statistics and charts presented in this paper may be readily reproduced
from \texttt{growcurves}. The parameters under DP priors are all sampled
in a conjugate fashion by marginalizing over the random measure, $F$,
to produce the P\'{o}lya urn scheme of \citet{blackwell1973}, under
which each cluster assignment indicator is sampled from a mixture of
existing clusters and a new cluster. To the extent that a new cluster
is selected, associated parameter locations are generated (and
subsequently resampled) from the posterior of the base distribution
under a single observation. [See \citet{PaddSaviJRSSA} for details.]

We employ the cross-validatory, log pseudo marginal likelihood (LPML)
leave-one-out fit statistic as described in \citet{congdon2005bayesian}
under importance resampling of the posterior distributions over model
parameters to estimate $f (y_{i}|\mathbf{y}_{-i},M_{r} )$,
where $M_{r}$ indexes our models where the leave-one-out property
induces a penalty for model complexity and helps to assess the
possibility for overfitting. We also include the $\mathrm{DIC}_{3}$ criterion of
\citet{CeleForbRobeTittrepl2006} that composes the marginal
(predictive) density $\widehat{f} (\mathbf{y} )$ to estimate
$f(\mathbf{y}|\theta)$ for composition of $pD$ which is more
appropriate for the (DP or DDP) mixture formulations that characterize
all of our models. The nonpenalized mean deviance, $\bar{\mathbf{D}}$,
is also utilized.

\section{Simulation study} \label{simsect}
\subsection{Data generation}
We generate data sets for simulation modeling from~(\ref{modddp}) by
allocating the first $132$ clients to the CBT and a remaining $168$ to
a nongroup therapy usual care (UC) condition. We employ $24$ modules
for our simulation. Each CBT client attends $4$ modules and each module
on average holds $22$ clients. The module attendance sequences, $\{
\mathbf{x}_{i}\}$, used to select columns of the client-indexed matrix
effects, $\{\bolds{\Delta}_{i}\}$, are next generated in an
open-enrollment manner by randomly selecting the starting module for
each CBT client in the block of $4$ modules to which they are assigned.
We set $\mathbf{x}_{i} = [\mathbf{1},\mathbf{0},\ldots,\mathbf
{0} ]$ for all UC clients (who, by design, do not attend group
therapy modules) as our hold-out or comparator module attendance
sequence for identification. Such a design instantiates partial
overlaps among the module attendance sequences for clients. The minimum
and maximum numbers of clients linked to modules were restricted to
$11$ and $26$, respectively, to conform to practical limitations on the
underlying structure for group therapy modules. We simulate up to three
repeated measures per client.

We simulate $4$ clusters of clients, where each cluster generates a $(q
= 3) \times((S=24)+1)$ set of effect locations, $\bolds{\Delta
}_{m}^{*}$, shared by all clients assigned to them. The $q = 3$ rows of
$\bolds{\Delta}_{m}^{*}$ capture up to second order (intercept,
linear, and quadratic) polynomial effects for each module. The effects
are generated in a vectorized fashion from a multivariate Gaussian with
the covariance formulation as outlined for the DDP base distribution
enumerated in Section \ref{base}. The module effects are generated from
a multivariate proper CAR prior under the assumption of adjacency for
successive modules with smoothing parameter $\rho= 0.7$. A covariance
matrix allowing for $q = 3$ polynomial orders of module random effects
is defined with
\[
\bolds{\Lambda}^{-1} = %
\left[\matrix{ 50 & -12 & 0.5
\vspace*{2pt}
\cr
-12 & 16 & -1.2 \vspace*{2pt}
\cr
0.5 & -1.2 & 0.12 }\right] ,
\]
where the diagonals encode the variance of the first through third
polynomial orders, respectively, for each of $(S = 24) \times1$
multivariate cluster effect locations. We formulate $\bolds{\Lambda
}^{-1}$ such that the first and second orders and the second and third
orders express negative correlations; for example, if the slope for the
effect trajectory of a given module expresses a negative trajectory,
then the quadratic term is positive and will tend to decelerate or bend
the curve back up. Once the effects are generated, clients are randomly
assigned to one of the $4$ clusters with equal probability. Each
cluster will hold both UC and CBT clients, though the module attendance
sequence for the UC clients is set to $0$'s such that their assigned
module effects do not contribute to the generation of the response
values. The model intercept, $\mu$, is set to $35$ and fixed effect
coefficients are set to $\beta= (-3,0.25,0,-2.5,0.25 )$ for
$d_{ij} = (t_{ij},t_{ij}^{2},T_{i},T_{i}t_{ij},T_{i}t_{ij}^2)$,
respectively, for each client, $i$, where $T_{i}$ is an indicator for
the treatment arm assigned to client $i$ ($T_{i}=1$ for CBT, $T_{i}=0$
for UC) and $t_{ij}$ denotes the $j=1,\ldots,3$ continuously-valued
time at which $y_{ij}$ was observed, taking on value $0, 3,$ or $6$
months. The $q \times(S+1)$ resultant set of random effects for client
$i, \bolds{\Delta}_{i}$, are multiplied with the $(S+1) \times1$ MM
link vector, $\mathbf{x}_{i}$, to produce $q \times1, \bolds{\theta
}_{\mathbf{x},i}$ matched to $\mathbf{z}_{ij} = (1,t_{ij},t_{ij}^2)$
for client-specific polynomial variation from the mean time trend
(which is captured in $\bolds{\beta}$). The model noise precision is
set to $\tau_{\varepsilon} = 0.1$.

\subsection{Data modeling}

\begin{figure}

\includegraphics{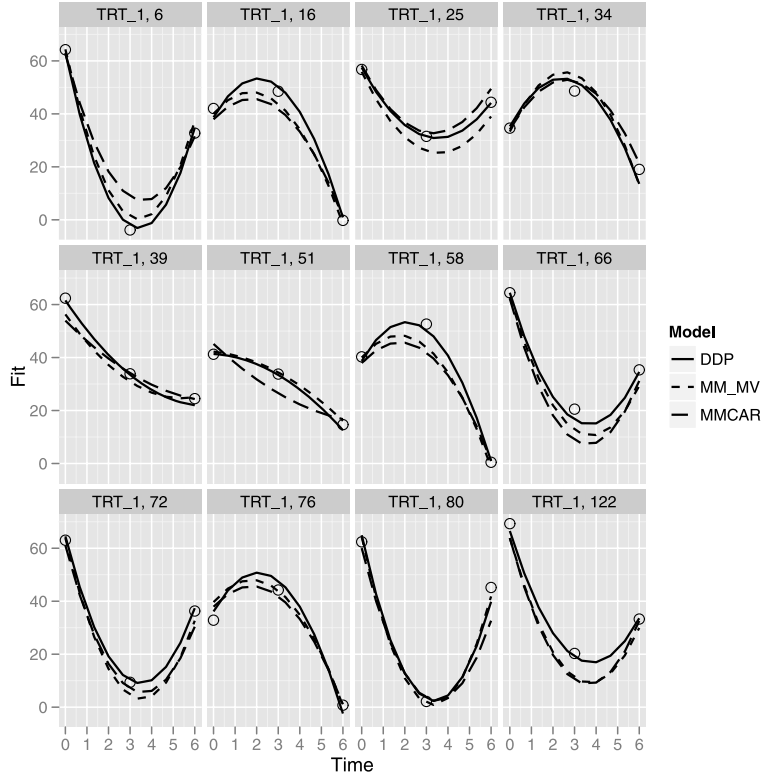}

\caption{Posterior mean client growth curves under semi-parametric
(MMCAR, MM\_MV) and nonparametric (DDP) MM models under simulated data.
Simulated data shown by circles.}
\label{gcsubjsim}
\end{figure}

Figure \ref{gcsubjsim} presents in-sample predicted growth curves for
randomly selected clients within each treatment arm along with actual
client data values. Client growth curves under the DDP model express
more adaptiveness to the data, both for $U$-shaped curves as expressed by
client $6$ and bell-shaped curves estimated for client $58$.
%

%

\begin{figure}

\includegraphics{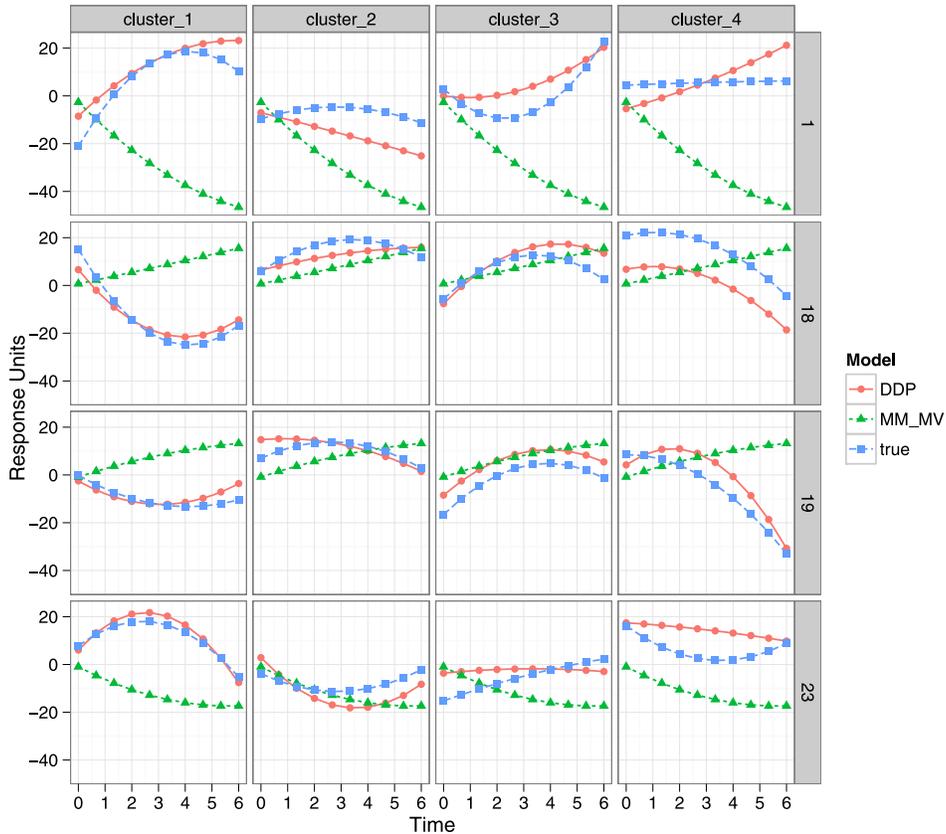}

\caption{Posterior mean client module effect trajectories from
simulated data for four clusters of clients, where the columns index
clusters and the rows represent randomly selected modules. The curves
are dimensioned in response units and represent the contribution of the
modules to the response.}\label{ddpsimeffcurves}
\end{figure}

Posterior mean values for the $3$ polynomial effect terms assigned to
each module are composed into module effect trajectories through time
in Figure~\ref{ddpsimeffcurves} comparing MM\_MV and DDP models for
each of the $4$ clusters (columns) and for $4$ randomly selected
modules. The posterior mean module effect trajectories estimated under
the DDP model track closer to the true trajectory shapes than do the
nonclient adaptive curves for MM\_MV.
%
%

\begin{figure}[t!]

\includegraphics{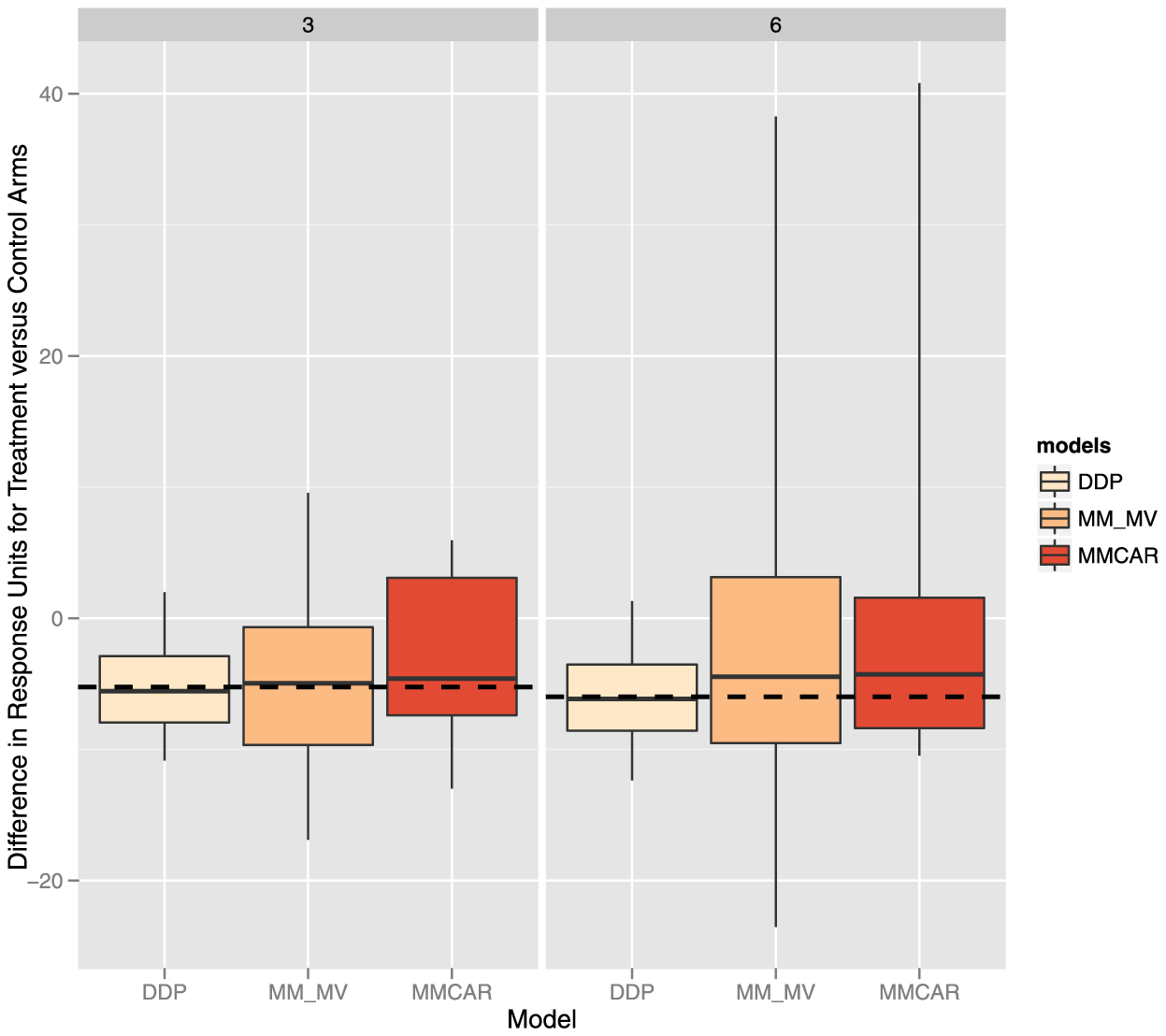}

\caption{Predictive margins for the treatment effect of CBT versus
usual care at $0$ (left panel), $3$~(middle panel), and $6$ (right
panel) months for MMCAR, MM\_MV, and DDP formulations under a Monte
Carlo simulation from a data-generating model where effects are indexed
by cluster of clients and modules. Segments reflect the $95\%$ credible
intervals and boxes represent the interquartile range of the marginal
posterior distribution. The dashed lines in each time period indicate
the true treatment effects over the simulated data sets.}
\label{trteff}
\end{figure}

We compose a small Monte Carlo simulation with $10$ iterations, where
each generates a data set with the above noted specifications.
Estimation is performed under our models for each generated data set
and the posterior draws for the fixed effects are concatenated across
iterations to examine performance of the 3 comparator formulations
under repeated sampling. Figure \ref{trteff} reveals the posterior
distribution over the $95\%$ credible intervals under each model
estimated using the predictive margins technique; see \citet
{LaneNeldanal1982}. We note that the DDP formulation expresses the
least uncertainty around the true values (represented by a dashed line
at each of the $3$ measurement months).

%

\subsubsection{Model fit statistics}
Model fit statistics, $\bar{D},$ $-\mathrm{LPML}$, and $\mathrm{DIC}_{3}$, are presented
in Table \ref{tabmodcompare}. One observes lower (better) values
across all $3$ statistics for the DDP than the other two comparator
models, while MM\_MV, employing multivariate module effects,
outperforms MMCAR parameterized with univariate module effects. In
particular, the leave-one-out LPML statistic strongly prefers the DDP
model. While the DDP is parameterized with client-by-module random
effects, the effective parameterization is reduced under the clustering
of clients. Nevertheless, the DDP would generally be expected to
express a higher number of effective parameters than the two additive
models, though the LPML performances do not indicate overfitting. The
polynomial construction for $\mathbf{z}_{ij}$ enforces smoothness in
the estimated fit as demonstrated in the client growth curves from
Figure \ref{gcsubjsim}, which also serves to mitigate the possibility
for overfitting. We performed additional simulations to explore
scenarios $48$ and $66$ modules that, on average, have $11$ and $8$
clients per module, respectively, with the same number of clients. The
relative model differences persist under $-\mathrm{LPML}$. The $-\mathrm{LPML}$
difference between DDP and MM\_MV is $158$ under $S = 24$ modules and
$149$ under $48$ modules and $251$ under $66$ modules.

%
\begin{table}[t!]
\caption{Simulation Study Model Fit Comparisons: $\bar{\mathbf{D}}$,
$-\mathrm{LPML},$ and $\mathrm{DIC}_{3}$ scores for model alternatives. Lower values
imply better performance}\label{tabmodcompare}
\begin{tabular*}{\textwidth}{@{\extracolsep{\fill}}lccc@{}}
\hline
\multicolumn{1}{@{}l}{\textbf{Model}} & $\bolds{\bar{D}}$ &
$\mathbf{-LPML}$ & $\mathbf{DIC_{3}}$\\
\hline
MMCAR & $5073$ & $2691$ & $5208$ \\
MM\_MV & $4905$ & $2592$ & $5034$ \\
DDP & $4607$ & $2434$ & $4715$ \\
\hline
\end{tabular*} 
\end{table}

\section{Application to group therapy data} \label{case}
We now return to the BRIGHT study for comparison of fit among our $3$
model formulations. We further focus on inference under the MM DDP
construction and examine heterogeneity with respect to module type in
BDI-II trajectories across disjoint clusters of clients. We recall our
parameterization of fixed effects for the BRIGHT study data, $\mathbf
{d}_{ij} = (T_{i}, t_{ij}, t^2_{ij},T_{i} t_{ij}, T_{i}
t^2_{ij} )$, where $T_{i}$ is an indicator for the treatment\vspace*{1pt} arm
assigned to client $i$ ($T_{i}=1$ for CBT, $T_{i}=0$ for UC) and
$t_{ij}$ denotes the continuously-valued time at which depressive
symptom score, $y_{ij},$ was observed. As before, set $\mathbf{z}_{ij}
= (1,t_{ij},t_{ij}^2)$.

We simplify and focus inference by composing posterior distributions
for module effects up to clusters of clients. The client clustering is
obtained from among posterior samples of client partitions using the
least squares algorithm of \citet{dahl2008}. The shapes, magnitudes,
and differences across the clusters express the range we see among
clients so that we do not lose generality with a focus at the cluster,
rather than client, level. The most populated $6$ clusters are employed
and contain $(88,51,24,23,20,19)$ clients, respectively, that together
hold $225$ out of $299$ total BRIGHT study clients. Roughly half of the
clients in the $6$ clusters are UC clients who do not attend any group
therapy modules. UC clients with mean client random effects, $\mathbf
{b}_{i}$, similar to those of a subset of CBT clients are expected to
co-cluster in posterior sampling such that the module effect values for
all clients in the cluster are assigned the module effect location
values for that cluster. This is an intuitive result where UC clients
who express similar idiosyncratic characteristics to co-clustered CBT
clients would be expected to similarly respond to CBT treatment were it
offered to them.

%
%
\begin{figure}

\includegraphics{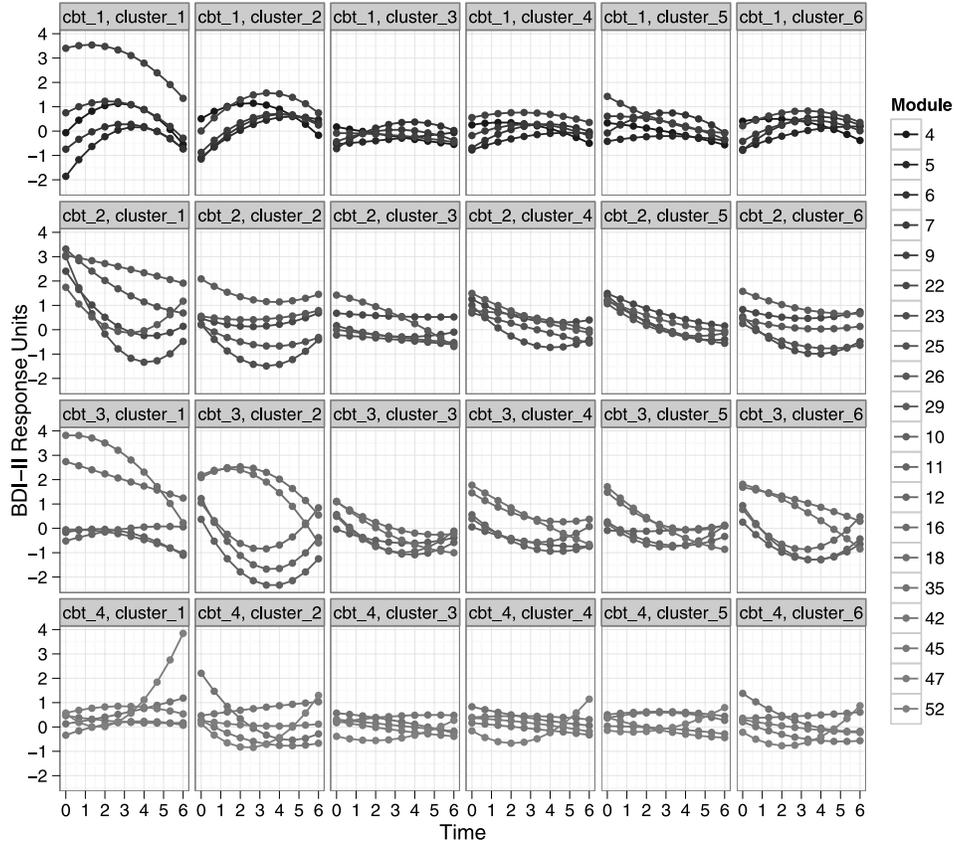}

\caption{Posterior mean client module effect trajectories of BDI-II
depressive symptom scores under the DDP model. Results are summarized
by averaging trajectories into client clusters. Each plot cell is
indexed by client cluster within each of $4$ disjoint CBT groups,
$(\mathrm{cbt}_{-1},\ldots,\mathrm{cbt}_{-4})$. The rows of plot cells
are indexed by CBT group and the columns by cluster. The largest $6$
clusters of clients are represented (in order of number of clients
contained in each). Each plot contains module effect trajectories for
randomly selected modules within each of the four CBT groups.}
\label{ddpeffcurves}
\end{figure}

Figure \ref{ddpeffcurves} renders module effect trajectories of the
BDI-II depressive symptom scores for randomly selected modules. Results
are summarized by averaging trajectories into client clusters, with the
largest six clusters shown across the columns, denoted by
$\mathrm
{cluster}\_1,\ldots,\mathrm{cluster}\_6$. Each client cluster's
trajectories are presented for each of the $4$ open-enrollment CBT
therapy groups along the rows within clusters, which are denoted by
$\mathrm{cbt}\_1,\ldots, \mathrm{cbt}\_4$ in the figure.
Large differences
are observed in shape and magnitude among modules, particularly for
client cluster $1$, whose trajectories for each of the four
open-enrollment groups are provided in the leftmost column of plot
cells of Figure \ref{ddpeffcurves}. The range of the curves expresses
clinically meaningful differences of 4--6 (BDI-II) points [\citet
{FuruAsse2010}]. Scanning the columns from left to right reveals a
marked attenuation in cluster responsiveness to the CBT intervention.
Member clients of clusters 4--6 express much less depressive symptom
sensitivity to participation in the modules and, therefore, one notes
much less differentiation in effect values among the modules for these clusters.
%

%
\begin{figure}

\includegraphics{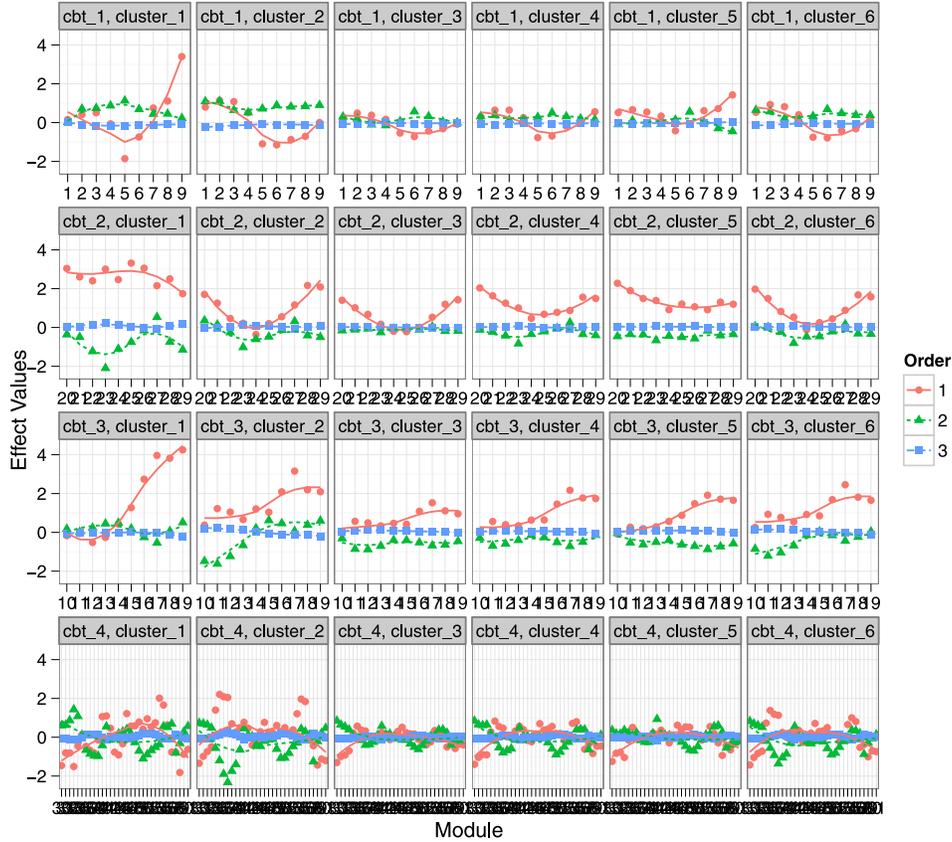}

\caption{Posterior mean client module intercept (order $1$), linear
(order $2$), and quadratic (order $3$) effects for each module within
the $4$ disjoint CBT groups of modules averaged to cluster for BRIGHT
case study under the DDP model. The rows of plot cells are indexed by
CBT group and the columns by cluster.}
\label{ddpeffs}
\end{figure}

Figure \ref{ddpeffs} provides additional insight from the DDP model for
examining the variation in module effects across clusters of clients
and how those effects vary over time. The figure shows module effect
trajectories disaggregated into the $q = 3$ posterior mean polynomial
effects from which they are each rendered across the $6$ clusters of
clients. The $3$ polynomial effect values are presented for all
modules, organized in the same cbt group-within-cluster format utilized
in Figure \ref{ddpeffcurves}. These polynomial parameters imply a
module effect trajectory with the order $1$ effect providing the
intercept, the order $2$ effect the slope and order $3$ a nonlinear
quadratic term. The resulting effects trajectory for a module would be
$U$-shaped if the order~$3$ term is positive. As we noted in Figure \ref
{ddpeffcurves}, there is notable variation in the effect of modules on
depressive symptoms across client clusters within each of the four CBT
therapy groups as we scan from left to right, particularly for cbt
groups 1--3; for example, the first two clusters of each CBT therapy
group, shown in the first two columns of Figure \ref{ddpeffs}, show
clinically meaningful variation in client outcomes.

Model fit statistics, presented in Table \ref{tab2}, reveal an improved fit for the DDP in comparison
to the other two models, however, unlike for the simulation results,
the MMCAR produces a better fit for the BRIGHT data than does MM\_MV.
These results indicate the importance of differences across clients in
responsiveness to modules. Within-sample predicted growth curves (not
shown) demonstrate a similar improvement as observed in Figure \ref
{gcsubjsim} in shape and orientation adaptability for the DDP as
compared to the other models to express better fit performance.
%
%
\begin{table}
\caption{BRIGHT Study Model Fit Comparisons: $\bar{D}$, $-\mathrm{LPML},$ and
$\mathrm{DIC}_{3}$ scores for model alternatives. Lower values imply better performance}\label{tab2}
\begin{tabular*}{\textwidth}{@{\extracolsep{\fill}}lccc@{}}
\hline
\multicolumn{1}{@{}l}{\textbf{Model}} & $\bolds{\bar{D}}$ &
$\mathbf{-LPML}$ & $\mathbf{DIC_{3}}$ \\
\hline
MMCAR & $5505$ & $2980$ & $5679$\\
MM\_MV & $5547$ & $2994$ & $5716$ \\
DDP & $5079$ & $2929$ & $5302$ \\
\hline
\end{tabular*}
\end{table}

We explore sensitivity of the clustering of clients to our prior
specification for the DP concentration parameter, $\alpha$, employed in
(\ref{Fprior}) for the MM DDP model by varying the shape and rate
hyperparameters, $ (a_{1},b_{1} )$, employed in the prior,
$\alpha\sim\mathcal{G}a (a_{1},b_{1} )$. We vary both
hyperparameters in combinations within a range of 1--4 for each,
producing a prior number of clusters from a minimum of $3$ to a maximum
of $18$. While our group therapy data application results show small
differences in the posterior numbers of clusters formed, the allocation
of clients to the most populous clusters is essentially unchanged, as
is our inference on client-module effects. Distributions for underlying
parameters are also essentially unchanged.

\section{Discussion} \label{discusssect}
Our MM DDP approach extends the ANOVA DDP construction of \citet
{DeIMlleRosnMacEan2004} to a multiple membership framework. The MM
DDP provides wide support on the space of distributions indexed by the
set of distinct multiple membership sequences through the borrowing of
strength in overlaps among expressed sequences. The formulation allows
one to examine whether element (e.g., module) effects vary across
different client trajectories and vice versa, allowing for one to learn
about differing response sensitivities among clients to treatment
elements, even for unobserved combinations of clients and treatment
elements. We compose a model base distribution to retain
straight-forward and efficient posterior sampling properties of the DP
while allowing flexibility for Gaussian covariance specifications to
parsimoniously parameterize dependence among module effects; in
particular, we illustrate adjacency-based formulations for the
covariance matrix of the Gaussian base measure in a fashion that
renders flexibility while retaining conjugacy.


Other alternatives to our MM DDP may be considered, such as the
hierarchical DP (HDP) [\citet{TehJordBealBleihier2006}] or the
nested DP (NDP) [\citet{rodriguez2008}], which both target a grouped
data structure with nested observations. These approaches, however,
do not anticipate a multiple membership construction where subgroups of
clients share connections to the same modules as does the MM DDP, which
indexes the collection of random measures, $\{F_{\mathbf{x}}\}$, by
multiple membership (attendance) sequence. While one may ignore the
multiple membership composition and employ either of the HDP and NDP,
they both perform posterior simulations in a nested, two-step, fashion
(for a two-level hierarchical formulation), while we see how the MM DDP
reduces to a DP that permits a simpler computational approach. Last,
neither the HDP or NDP allow inference on unobserved module attendance
sequences as does the MM DDP.

The usefulness of our approach may be limited for data with decreasing
overlaps among the treatment element (e.g., client module attendance)
sequences, $\{\mathbf{x}_{i}\}$, as this would restrict the ability for
the data to borrow strength in the estimation of the collection of
random distributions, $\{F_{\mathbf{x}}\}$. In one direction where
clients perfectly overlap into disjoint groupings of client-modules for
CBT studies, the MM DDP reduces to the ANOVA DDP. In the other
direction, however, where clients express progressively less overlaps
in modules attended, estimability may be compromised. In practice,
resource limitations in the total number of modules offered for typical
open-enrollment group therapy studies tend to produce a sufficient
level of overlaps of clients on each module for estimation.

Software implementing the MM DDP is available for the R statistical
software [\citet{rdctr}] in a package called \texttt{growcurves} [\citet
{SaviPaddgrow2011}]. All of the methods, fit statistics, and charts
presented in this paper may be reproduced from \texttt{growcurves}.

%
%

%



\printaddresses

\end{document}